\documentclass[a4paper,11pt]{article}
\usepackage[toc,page]{appendix}
\usepackage{graphicx}
\usepackage{colortbl}
\usepackage{amsmath}
\usepackage{subfigure}
\usepackage{mathtools}

\DeclarePairedDelimiterX\braket[2]{\langle}{\rangle}{#1 \delimsize\vert #2}
\usepackage[utf8]{inputenc}
\usepackage{float}
\usepackage[normalem]{ulem}
\usepackage{epstopdf}
\usepackage{amsfonts,amsmath,amssymb}
\usepackage{hyperref}

\newcommand{\f}[2]{\frac{#1}{#2}}
\usepackage{epsfig,multicol,bbm}
\usepackage{color}
\usepackage[dvipsnames]{xcolor}

\definecolor{darkblue}{rgb}{0.0, 0.0, 0.55}
\definecolor{grey}{rgb}{0.57, 0.64, 0.69}
\definecolor{lightbrown}{rgb}{0.71, 0.4, 0.11}

\date{}
\newcommand{\be}{\begin{equation}}
\newcommand{\ee}{\end{equation}}

\def\ndelta{\delta\hspace{-0.50em}\slash\hspace{-0.05em} }

\newcommand\fverb{\setbox\pippobox=\hbox\bgroup\verb}
\newcommand\fverbit{\egroup\item[\fbox{\unhbox\pippobox}]}

\newbox\pippobox
\begin{document}
\title{\bf Boundary Conditions of Warped AdS$_3$ in New Massive Gravity}
\author{S. N. Sajadi\thanks{Electronic address: naseh.sajadi@gmail.com}\,,\,
Supakchai Ponglertsakul\thanks{Electronic address: supakchai.p@gmail.com}\,,\,
\\
\small Strong Gravity Group, Department of Physics, Faculty of Science, Silpakorn University,\\ \small Nakhon Pathom 73000, Thailand\\
}
\maketitle
\begin{abstract}
To satisfy the Cardy formula for Warped AdS$_3$ (WAdS$_{3}$) solutions in a quadratic ensemble, a specific set of boundary conditions has been proposed in \cite{Aggarwal:2020igb}. In this paper, these boundary conditions have been investigated in the three-dimensional new massive gravity (NMG) framework. The associated solution space, asymptotic symmetries, and charge algebra have been extracted. It has been shown that the surface charges are finite, but not integrable, and the integrability of the charges is obtained after restricting to a sub-sector of the original solution space. Finally, we have proved that the Cardy formula reproduces the thermodynamic entropy of a warped BTZ black hole.

\end{abstract}

\maketitle
\section{Introduction}
Three-dimensional space-time provides a unique aspect for studying gravity because of its simplicity. Einstein’s gravity in three dimensions with a negative cosmological constant admits black hole solutions, but does not have any bulk propagating degrees of freedom \cite{Banados:1992wn,Banados:1992gq}.
Including the topological, Chern-Simons term for the Christoffel connection gives an odd parity, unitary theory in which its linearization around the maximally symmetric background gives a propagating massive graviton \cite{TMG1,Grumiller:2008qz}. Along this way, many theories are found to admit asymptotically anti de-Sitter (AdS) black hole solutions. However, they are not a proper testing ground for studying AdS/CFT correspondence. The trouble is that in order to have a massive graviton with positive energy in the bulk, it requires a negative central charge within the dual theory. This makes the dual conformal field theory (CFT) non-unitary. The first attempt to solve this problem is to study the theory at a critical point. But this leads to the non-unitary logarithmic CFT \cite{Grumiller:2010rm}. Some efforts have been made to modify the theory proposed in which the $R^{3}$ term is added to the action, and also the Born-Infeld type action, but the bulk-boundary clash remains persistent \cite{Sinha:2010ai,Gullu:2014gza}. An alternative theory introduced by Bergshoeff, et. al \cite{MMG} which eludes this problem is the Minimal Massive Gravity (MMG). The MMG has the same local structure as topologically massive gravity (TMG) \cite{Deser:1982vy} and fulfills the field equations of TMG with an asymmetric, rank-two tensor containing up to second derivatives of the metric ($J_{\mu \nu}$) \cite{MMG}. The remarkable feature of the MMG is that it is free from negative energy bulk modes and also avoids the bulk-boundary disagreement. 

On the other hand, New massive gravity (NMG) is a three-dimensional gravity theory with higher derivative terms that admits several vacua, including Warped AdS$_3$ (WAdS$_3$)\footnote{In some research communities, Warped AdS$_3$ and Squashed AdS$_3$ may be used interchangeably. This is because they have the same symmetry group $SL(2,R)\times U(1)$. However, they are basically different. Warped AdS$_3$ has a non-constant warping factor while for the squashed AdS$_3$ has a constant one. Thus, the squashed AdS$_3$ can be considered as a specific case of Warped AdS$_3$ \cite{Anninos:2008fx}.  }\cite{Bergshoeff:2009hq,Bergshoeff:2009aq}. This theory is a parity-even covariant theory of gravity which at the linearized level reduces to a massive spin-two Fierz-Pauli theory \cite{Fierz:1939ix}. When linearized around AdS$_3$ vacua, NMG exhibits features that are the same as Topologically Massive Gravity (TMG) \cite{Deser:1982vy}. More specifically, at the chiral point, the central charge of the dual two-dimensional CFT vanishes, and the asymptotic boundary conditions are relaxed to the Brown-Henneaux boundary conditions \cite{Brown:1986nw}. The pivotal role of AdS/CFT in three-dimensional spacetime is investigated by Brown and Henneaux \cite{Brown:1986nw}. They show that the symmetry algebra of asymptotically AdS$_3$ spaces is generated by two copies of the Virasoro algebra with non-vanishing central charge \cite{Brown:1986nw}. Compere, Song, and Strominger (CSS) \cite{Compere:2013bya} have demonstrated a family of specific alternative boundary conditions in which the asymptotic symmetry algebra of a $3D$ theory turns out to be consisted of a direct product of a Virasoro and $u(1)$ Kac-Moody algebras, which are symmetries of the 2-dimensional Warped-CFTs.
In \cite{Ciambelli:2020shy,Setare:2021ugr}, TMG and general minimal massive gravity (GMMG) are studied under the CSS boundary conditions. In \cite{Aggarwal:2020igb}, a new set of boundary
conditions in three-dimensional topologically massive gravity is introduced so that the dual field theory is a Warped Conformal Field theory (WCFT) in the quadratic ensemble. In \cite{Sajadi:2022hxo}, by considering the boundary conditions proposed in \cite{Aggarwal:2020igb}, the solution space, asymptotic symmetries, and charge algebra have been studied within the framework of general massive gravity (GMG).

In this work, along the line of works \cite{Aggarwal:2020igb,Sajadi:2022hxo}, we introduce the boundary conditions and study their consequences. These include solution space, asymptotic symmetries, and charge algebra in the framework of the three-dimensional new massive gravity theory. In \cite{Donnay:2015iia}, the same analyses have been done with the boundary conditions that have been obtained from WAdS$_{3}$ black holes. It belongs to a phase space whose asymptotic symmetry algebra is given by that of a WCFT in a canonical ensemble. While the boundary conditions considered here have been obtained from the WBTZ line element, which is dual to WCFT in the quadratic ensemble. A motivation for introducing these new boundary conditions is that the warped Cardy formula is well defined only for WCFTs dual to WBTZ coordinates \cite{Aggarwal:2020igb}. 

The paper is organized as follows: In Section \ref{sec2}, after introducing the NMG theory and boundary conditions, we impose the field equations to determine the solution space. In section \ref{secc3}, we compute the asymptotic residual vectors preserving the gauge conditions and their corresponding surface charges in the NMG. 
In section \ref{secc4}, we compute the bulk thermodynamic entropy and compare it with the warped Cardy formula,  showing that they match.
We provide some conclusions in section \ref{secc5}.

\section{NMG under the new boundary conditions}\label{sec2}

One of the famous theories of gravity in three dimensions is the new massive gravity. This model at the linearized level is equivalent to the three-dimensional Fierz-Pauli action for a massive spin$-2$ field. Furthermore, the NMG preserves parity symmetry, which is not the case for topologically massive gravity (TMG).
The action of the NMG is presented as follows \cite{Bergshoeff:2009hq}
\begin{equation}\label{action}
S_{NMG}=\dfrac{1}{8\pi G}\int d^{3}x\sqrt{-g}\left[\mathcal{R}-2\lambda -\dfrac{1}{\Xi^2}\left(\mathcal{R}^{\mu \nu}\mathcal{R}_{\mu \nu}-\dfrac{3}{8}\mathcal{R}^{2}\right)\right],
\end{equation}
where $\lambda$ and $\Xi$ are the cosmological constant and the parameter of NMG, respectively. The relative coefficient $3/8$ between the two quadratic terms is necessary for the theory to be free of ghosts. By a variation of the Lagrangian, one can obtain
\begin{equation}\label{eq1}
\mathcal{E}_{\mu \nu}=\mathcal{G}_{\mu \nu}+\lambda g_{\mu \nu}-\dfrac{1}{2\Xi^{2}}\mathcal{K}_{\mu \nu}=0, \,
\end{equation}
where
\begin{equation}\label{eq2}
\mathcal{K}_{\mu \nu}=-\dfrac{1}{2}\nabla^{2}\mathcal{R} g_{\mu \nu}-\dfrac{1}{2}\nabla_{\mu}\nabla_{\nu} \mathcal{R}+2\nabla^{2}\mathcal{R}_{\mu \nu}+4 \mathcal{R}_{\mu \alpha \nu \beta}\mathcal{R}^{\alpha \beta}-
\dfrac{3}{2}\mathcal{R} \mathcal{R}_{\mu \nu}-\mathcal{R}_{\alpha \beta}\mathcal{R}^{\alpha \beta}g_{\mu \nu}+\dfrac{3}{8}\mathcal{R}^{2}g_{\mu \nu},
\end{equation}
and $\mathcal{G}_{\mu \nu}$ is the Einstein tensor. Notice that $\mathcal{K}_{\mu\nu}$ contains fourth-order derivatives of the metric tensor. 
We consider the line element in the Fefferman-Graham gauge with the coordinate $ x^{\mu}=(r, x^{+},x^{-}) $ (such that $x^{\pm}=t/L\pm \phi$) as follows
\begin{equation}\label{metric}
ds^2=\dfrac{L^2}{r^2}dr^2+\gamma_{\alpha \beta}(r, x)dx^{\alpha}dx^{\beta},
\end{equation}
with gauge conditions 
\begin{equation}\label{gaugcond}
g_{rr}=\dfrac{L^2}{r^2},\;\;\;\;\; g_{r a}=0,
\end{equation}
here $L$ is a quantity with dimension of length.
Comparing with the WBTZ metric (see Appendix \ref{app:A}), we solve the field equations by considering the following boundary conditions \cite{Aggarwal:2020igb},\cite{Sajadi:2022hxo}
\begin{equation}\label{boundcond}
 \gamma_{+ +}=\mathcal{O}(r^4),\;\;\;\gamma_{+ -}=\mathcal{O}(r^2),\;\;\; \gamma_{- -}=\mathcal{O}(1).
\end{equation}
Therefore, the field equation \eqref{eq1} gives us the coefficients of expansion as follows 
\begin{align}
\gamma_{+ +}&=j_{++}r^{4}+h(x^{+})r^2+f_{++}(x^{+}) \nonumber\\
&~~~+\dfrac{h(x^{+})\left(\Xi^2 L^2-\dfrac{17}{2}\right)\left[j_{++}f_{++}(x^{+})(\Xi^2 L^2+2)+h^{2}(x^{+})\left(\dfrac{17}{8}-\dfrac{\Xi^2 L^2}{4}\right)\right]}{j_{++}^{2}r^2(2\Xi^4 L^4+29\Xi^2 L^2+50)}\nonumber\\
&~~~+\dfrac{\left(\Xi^2 L^2-\dfrac{17}{2}\right)^{2}\left[(\Xi^2 L^2+2)j_{++}h(x^{+})-\dfrac{1}{4}h^{2}(x^{+})\left(\Xi^2 L^2-\dfrac{17}{2}\right)\right]^{2}}{4r^4 j_{++}^{3}(\Xi^2 L^2+2)^{2}\left(\Xi^2 L^2+\dfrac{25}{2}\right)^{2}},\\
\gamma_{+ -}&=\varsigma_{+ -} r^2+\dfrac{(2\Xi^2 L^2-17)h(x^{+})\varsigma_{+ -}}{4(\Xi^2 L^2+2)j_{++}}\nonumber\\
&~~~+\dfrac{\varsigma_{+-}\left(\Xi^2 L^2-\dfrac{17}{2}\right)\left[(\Xi^2 L^2+2)j_{++}f_{++}(x^{+})-\dfrac{1}{4}h^{2}(x^{+})\left(\Xi^2 L^2-\dfrac{17}{2}\right)\right]}{j_{++}^{2}(2\Xi^4 L^4+29\Xi^2 L^2+50)r^2},\\
\gamma_{- -}&=\dfrac{(2\Xi^2 L^2-17)\varsigma_{+ -}^{2}}{2(\Xi^2 L^2+2)j_{++}}.
\end{align}
These expansion coefficients satisfy the field equation \eqref{eq1} when the following holds
\begin{equation}
\lambda=\dfrac{\Xi^4 L^4-32\Xi^2 L^2+16}{21\Xi^2 L^4}.
\end{equation}
In order to have a negative cosmological constant, it requires
\begin{equation}
4(4-\sqrt{15})<\Xi^2 L^2 <4(4+\sqrt{15}),
\end{equation}
where we assume that $\Xi>0$ and $L>0$. The Ricci scalar of \eqref{metric} is given by
\begin{equation}
\mathcal{R}=\dfrac{4(\Xi^2 L^2-40)}{21L^2}.
\end{equation}
Since in three-dimensional AdS spacetime, the Ricci scalar must be $6\lambda$. This provides an additional constraint, that is, $\Xi^2L^2=4$ and $12$. Moreover, the Ricci scalar has the same sign as $\lambda$. The solution space is characterized by four quantities: two constants $j_{++}$, $\varsigma_{+-}$, and two chiral functions $h(x_{+})$, $f_{++}(x_{+})$. Remark that our solution space is larger than that in \cite{Aggarwal:2020igb}. This is because $\varsigma_{+-}$ is not fixed in our case.
The line element \eqref{metric} has two interesting limits. The first one is the null warped limit obtained
by setting $2\Xi^2 L^2=17$ while keeping $j_{++}$ arbitrary.
In this case, the cosmological constant becomes 
\begin{equation}
\lambda =-\dfrac{1+4\Xi^2 L^2}{4\Xi^2 L^4},
\end{equation}
and the line element leads to
\begin{equation}
ds^2=\dfrac{L^2}{r^2}dr^2+(j_{++}r^{4}+h(x^{+})r^2+f_{++}(x^{+}))dx^{+2}+\varsigma_{+-}r^2dx^{+}dx^{-}. \label{firstmetric}
\end{equation}
This metric is not a solution for the pure Einstein equation, because the NMG part has non-vanishing components ($\mathcal{K}^{+}_{-}=-\frac{136j_{++}r^2}{\varsigma_{+-}L^4}$). 

The second limit is the CSS limit. It is performed as $j_{++}\to 0$ 
and $2\Xi^2 L^2\to 17$. In this case, the corresponding line element is \begin{align}\label{eqcss1}
ds^{2}&=\dfrac{L^2}{r^2}dr^2+\left[h(x^{+})r^2+f_{++}(x^{+})+\dfrac{h(x^{+})\Delta (84f_{++}(x^{+})+\Delta h^{2}(x^{+}))}{3528r^2}\right]dx^{+2}\nonumber\\
&+\dfrac{\Delta \varsigma_{+-}^{2}}{21}dx^{-2}+\left[\varsigma_{+-}r^2+\dfrac{h(x^{+})\varsigma_{+-}\Delta}{42}+\dfrac{\varsigma_{+-}\Delta(42f_{++}(x^{+})-h^{2}(x^{+})\Delta)}{1764r^2}\right]dx^{+}dx^{-},
\end{align}
where
\begin{equation}
\Delta=\dfrac{2\Xi^2 L^2-17}{j_{++}}=constant.
\end{equation}
In this case, the metric above \eqref{eqcss1} is a solution of Einstein gravity. By writing the WBTZ black holes (see Appendix \ref{app:A}) (\ref{WBTZsol}) in the Fefferman-Graham gauge and including them in the boundary conditions \eqref{boundcond}, one gets
\begin{align}
j_{++}&=\dfrac{2\Xi^2 L^2-17}{168GL(ML-J)},\\
\varsigma_{+-} &={-}\dfrac{2}{21}(2+\Xi^2 L^2),\\
\gamma_{--}&=\dfrac{4GL(2+\Xi^2 L^2)(LM-J)}{21},\\
f_{++}(x^{+})&=\dfrac{GL (ML+J)(25+2\Xi^2 L^2)}{21},\\
h(x^{+})&=0.
\end{align}
where $G$ is the Newtonian constant and $J$ is a rotation parameter.

\section{Symmetries and charges}\label{secc3}
The residual gauge diffeomorphisms are those diffeomorphisms that satisfy the gauge conditions \eqref{gaugcond} as follows:
\begin{equation}
\mathcal{L}_{\xi}g_{rr}=0,\;\;\;\;\; \mathcal{L}_{\xi}g_{ra}=0,
\end{equation}
where $\xi$ is a vector and $\mathcal{L}$ denotes the Lie derivative.
The solution of these equations is
\begin{equation}\label{eqqkil}
\xi =\xi^{\mu}\partial_{\mu}=\xi^{r}\partial_{r}+\xi^{+}\partial_{+}+\xi^{-}\partial_{-},
\end{equation}
with
\footnotesize
\begin{align}
\xi^{r}&=r\eta(x^{+}),\nonumber \\
\xi^{+}&=\epsilon(x^+) \nonumber\\
&{\small-\dfrac{4(\Xi^2L^2 +2)(2\Xi^2L^2 +25)L^2(2\Xi^2 L^2-17)j_{++}\eta^{\prime}}{(2184\Xi^2L^2-336\Xi^4L^4+5712)j_{++}f_{++}(x^{+})+21(2\Xi^2L^2-17)^2h^{2}(x^{+})+336(\Xi^2L^2+2)(2\Xi^2L^2+25)r^4j_{++}^{2}}},\nonumber\\
\xi^{-}&=\sigma(x^+) \nonumber\\
&+\dfrac{(\Xi^2L^2+2)\left(\Xi^2L^2+\dfrac{25}{2}\right)L^2j_{++}\eta^{\prime}(\Xi^2L^2+2)r^2j_{++}+\left(\dfrac{\Xi^2L^2}{4}-\dfrac{17}{8}\right)h(x^{+})}{21\varsigma_{+-}\left[\dfrac{h^{2}(x^{+})}{8}\left(\Xi^2L^2-\dfrac{17}{2}\right)^{2}+(\Xi^2L^2+2)j_{++}\left(f_{++}(x^{+})\left(\dfrac{17}{4}-\dfrac{\Xi^2L^2}{2}\right)+\left(\Xi^2L^2+\dfrac{25}{2}\right)j_{++}r^4\right)\right]}.
\end{align}
\normalsize
In these expressions, $\sigma(x^{+})$ and $\epsilon(x^{+})$ are field-independent arbitrary chiral functions which are supertranslation in $x^{-}$ direction and $x^{+}$ direction, respectively. As we know, $\xi$ is not a Killing vector. Therefore, the variation of a metric (\ref{metric}) along the $\xi$ direction, we find the variation of solution space
\begin{equation}
\mathcal{L}_{\xi}g_{\mu \nu}dx^{\mu}dx^{\nu}=\dfrac{L^2}{r^2}dr^2+\delta_{\xi}\gamma_{a b}(r,x)dx^{a}dx^{b},
\end{equation}
where
\begin{align}
\delta_{\xi}\gamma_{++}&={\mathcal{L}_{\xi}} \gamma_{++}, \\
\delta_{\xi}j_{++}&=2j_{++}(\epsilon^{\prime}+2\eta),\label{eqdelta30}\\
\delta_{\xi}h &= 2h\eta +2\varsigma_{+-}\sigma^{\prime}+\epsilon h^{\prime}+2h\epsilon^{\prime},\\
\delta_{\xi}f_{++}&=\epsilon f_{++}^{\prime}+2f_{++}\epsilon^{\prime}+\dfrac{(2\Xi^2L^2-17)h\varsigma_{+-}\sigma^{\prime}}{2j_{++}(\Xi^2L^2+2)}+\dfrac{(2\Xi^4L^6+29\Xi^2L^4+50L^2)\eta^{\prime\prime}}{42(\Xi^2L^2+2)},
\end{align}
and
\begin{align}\label{eqq20}
\delta_{\xi}\gamma_{+-}={\mathcal{L}_{\xi}}\gamma_{+-}
\;\;\;\to\;\;\;\delta_{\xi}\varsigma_{+-}=\varsigma_{+-}(2\eta+\epsilon^{\prime}).
\end{align}
Here $\eta$ is a superscaling chiral function and prime denotes the derivative with respect to $x^{+}$.  By requiring $j_{++}$ to be constant, from \eqref{eqdelta30} one can get
\begin{equation}\label{eqq34}
\eta=-\dfrac{1}{2}\epsilon^{\prime}+\eta_{0},
\end{equation}
with arbitrary constant $\eta_0$. Therefore, \eqref{eqdelta30} becomes
\begin{equation}
\delta_{\xi}j_{++}=4j_{++}\eta_{0}.
\end{equation}
If we assume $\eta_{0}=0$, then we obtain
\begin{equation}
\delta_{\xi}j_{++}=\delta_{\xi}\varsigma_{+-}=0.
\end{equation}
This means that $j_{++}$ and $\varsigma_{+-}$ are fixed along the residual orbits. Finally, we find the full residual variation of the solution space as:
\begin{align}
\delta_{\xi}j_{++}&=0,\\
\delta_{\xi}h&=(h\epsilon)^{\prime}+2\varsigma_{+-}\sigma^{\prime},\label{eqqq34}\\
\delta_{\xi}f_{++}&=\epsilon f_{++}^{\prime}+2f_{++}\epsilon^{\prime}+\dfrac{(2\Xi^2L^2-17)h\varsigma_{+-}\sigma^{\prime}}{2j_{++}(\Xi^2L^2+2)}-\dfrac{(2\Xi^4L^6+29\Xi^2L^4+50L^2)\epsilon^{\prime\prime\prime}}{84(\Xi^2L^2+2)}.\label{eqqq35}
\end{align}
From the last term of \eqref{eqqq34} and \eqref{eqqq35}, one can see that $h$ and $f_{++}$ are $u(1)$ and Virasoro currents, respectively.
The general symmetries using \eqref{eqq34} are as follows:
\footnotesize
\begin{align}\label{eqqxi}
\xi^{r}&=-\dfrac{1}{2}r\epsilon^{\prime},\nonumber \\
\xi^{+}&=\epsilon +\dfrac{4(\Xi^2L^2 +2)(2\Xi^2L^2 +25)L^2(2\Xi^2 L^2-17)j_{++}\epsilon^{\prime\prime}}{2(2184\Xi^2L^2-336\Xi^4L^4+5712)j_{++}f_{++}(x^{+})+21(2\Xi^2L^2-17)^2h^{2}(x^{+})+336(\Xi^2L^2+2)(2\Xi^2L^2+25)r^4j_{++}^{2}},\nonumber\\
\xi^{-}&=\sigma- \dfrac{(\Xi^2L^2+2)\left(\Xi^2L^2+\dfrac{25}{2}\right)L^2j_{++}\epsilon^{\prime\prime}\left[(\Xi^2L^2+2)r^2j_{++}+\left(\dfrac{\Xi^2L^2}{4}-\dfrac{17}{8}\right)h(x^{+})\right]}{42\varsigma_{+-}\left[\dfrac{h^{2}(x^{+})}{8}\left(\Xi^2L^2-\dfrac{17}{2}\right)^{2}+(\Xi^2L^2+2)j_{++}\left(f_{++}(x^{+})\left(\dfrac{17}{4}-\dfrac{\Xi^2L^2}{2}\right)+\left(\Xi^2L^2+\dfrac{25}{2}\right)j_{++}r^4\right)\right]},
\end{align}
\normalsize
which depend on two arbitrary chiral functions $\epsilon(x^{+})$ (generates the usual Witt algebra) and $\sigma(x^{+})$ (generates an abelian algebra). These symmetry generators are one-to-one correspondence with the functions on the solution space $f_{++}(x^{+})$ and $h(x^{+})$. Therefore, the total asymptotic symmetry algebra is a direct sum of a Witt and a $u(1)$ algebra.

\subsection{Charges and Algebra}

By employing the Abbott–Deser–Tekin (ADT) method, the authors of \cite{Nam:2010ub},\cite{Devecioglu:2010sf},\cite{Ghodrati:2016vvf}, obtain expressions for the surface charges in the NMG theory. The expressions are given as follows:
\begin{align}\label{eq10}
\ndelta Q^{a}(\bar{\xi})=\int_{\Sigma}dS_{i}F^{a i}(g,h),
\end{align}
with
\begin{small}
\begin{align}
F_{E}^{a i}(\bar{\xi})=&\bar{\xi}_{b}\bar{\nabla}^{a}h^{i b}-\bar{\xi}_{b}\bar{\nabla}^{i}h^{a b}+\bar{\xi}^{a}\bar{\nabla}^{i}h-\bar{\xi}^{i}\bar{\nabla}^{a}h+h^{a b}\bar{\nabla}^{i}\bar{\xi}_{b}-h^{i b}\bar{\nabla}^{a}\bar{\xi}_{b}
 +\bar{\xi}^{i}\bar{\nabla}_{b}h^{a b}
 -\bar{\xi}^{a}\bar{\nabla}_{b}h^{i b}+h\bar{\nabla}^{a}\bar{\xi}^{i}\nonumber\\
F^{a b}_{\mathcal{R}^{2}}(\bar{\xi})=&2\mathcal{R}F^{a b}_{E}(\bar{\xi})+4\bar{\xi}^{[a}\nabla^{b]}\delta \mathcal{R}+2\delta \mathcal{R}\nabla^{[a}\bar{\xi}^{b]}-2\bar{\xi}^{[a}h^{b]\alpha}\nabla_{\alpha}\mathcal{R}\nonumber\\
F^{a b}_{\mathcal{R}_{2}}(\bar{\xi})=&\nabla^{2}F^{a b}_{E}+\dfrac{1}{2}F^{a b}_{\mathcal{R}^{2}}-2F_{E}^{\alpha[a}\mathcal{R}^{b]}_{\alpha}-2\nabla^{\alpha}\bar{\xi}^{\beta}\nabla_{\alpha}\nabla^{[a}h^{b]}_{\beta}-
4\bar{\xi}^{\alpha}\mathcal{R}_{\alpha \beta}\nabla^{[a}h^{b]\beta}-\mathcal{R}h_{\alpha}^{[a}\nabla^{b]}\bar{\xi}^{\alpha}+
2\bar{\xi}^{[a}\mathcal{R}^{b]}_{\alpha}\nabla_{\beta}h^{\alpha \beta}\nonumber\\
&+2\bar{\xi}_{\alpha}\mathcal{R}^{\alpha[a}\nabla_{\beta}h^{b]\beta}+2\bar{\xi}^{\alpha}h^{\beta[a}\nabla_{\beta}\mathcal{R}^{b]}_{\alpha}
+2h^{\alpha \beta}\bar{\xi}^{[a}\nabla_{\alpha}\mathcal{R}^{b]}_{\beta}-
(\delta \mathcal{R}+2\mathcal{R}^{\alpha \beta}h_{\alpha \beta})\nabla^{[a}\bar{\xi}^{b]}
-3\bar{\xi}^{\alpha}\mathcal{R}^{[a}_{\alpha}\nabla^{b]}h\nonumber\\
&-\bar{\xi}^{[a}\mathcal{R}^{b]\alpha}
\nabla_{\alpha}h,
\end{align}
\end{small}
where $\delta \mathcal{R}=-\mathcal{R}^{\alpha \beta}h_{\alpha \beta}+\nabla^{\alpha}\nabla^{\beta}h_{\alpha \beta}-\nabla^{2}h$, $\eta^{\nu}=\epsilon^{\nu \rho \sigma}\bar{\nabla}_{\rho}\bar{\xi}_{\sigma}$ and $h=\delta g_{\mu \nu}(\delta \alpha,\alpha)=\frac{\partial g_{\mu \nu}}{\partial \alpha} \delta \alpha$. We start with the $u(1)$ sector (for the vector $\underline{\sigma}=\sigma(x^{+})\partial_{-}$ and $\epsilon(x^{+})=0$) and compute the NMG surface charges. Thus, we have
{\begin{align}
{\ndelta Q_{\underline{\sigma}}}={-}\dfrac{2\sqrt{42}(2\Xi^2L^2-17)\varsigma_{+-}}{21\Xi^2L^3\sqrt{\Xi^2L^2+2}j_{++}^{2}}\int_{0}^{2\pi} d\phi \sigma(x^{+})[2j_{++}\delta h+(-h+2\varsigma_{+-})\delta j_{++}-4j_{++}\delta \varsigma_{+-}].
\end{align}}
The Virasoro charge is (for vector $\epsilon(x^+)\neq0$)
\begin{small}
\begin{align}
{\ndelta Q_{\underline{\epsilon}}}=&\dfrac{2\sqrt{42}}{1323\Xi^2L^3\sqrt{\Xi^2L^2+2}(2\Xi^2L^2+{25})j_{++}^{2}\varsigma_{+-}}\int_{0}^{2\pi}d\phi \Big[-{2L^2}(2\Xi^2L^2+{25})(2\Xi^2L^2-{17})(\Xi^2L^2+2) \varsigma_{+-}\nonumber\\
&j_{++}\delta j_{++}\epsilon^{\prime\prime}-189(2\Xi^2L^2-{17})j_{++}\varsigma_{+-}h\delta h(2\Xi^2L^2+{11})\epsilon+ 84(\xi^2L^2+2)^{2}\epsilon\varsigma_{+-}j_{++}^{2}\delta f_{++} +2(2\Xi^2L^2-{17})\nonumber\\
&\varsigma_{+-}\delta j_{++}(-(\Xi^2L^2+2)(2\Xi^2L^2+{25})L^2j_{++}\epsilon^{\prime\prime}+
 {63}(2\Xi^2L^2+{11})\epsilon h^{2}- 21(2\Xi^2L^2+{25})\epsilon\varsigma_{+-}h)+63 j_{++}\delta\varsigma_{+-}\nonumber\\
 &(2\Xi^2L^2+{25}) (-(\Xi^2L^2+2)L^2j_{++}\epsilon^{\prime\prime}+2(2\Xi^2L^2-{17})\epsilon h\varsigma_{+-})\Big],
\end{align}
\end{small}
where we have used $\underline{\epsilon}=\xi$, with $\xi^{r}$,  $\xi^{+}$ and  $\xi^{-}$ provided in \eqref{eqqxi} and $\sigma=0$. Here, we first obtain the surface charge for fixed ($r, x^{+}$) and then compute the charge at the surface fixed ($r, x^{-}$). After that, we combine the results from the two surfaces and take $r \to \infty$. Clearly, these charge algebras are non-integrable. This is due to the fact that there are propagating degrees of freedom. There are two possibilities to find integrable charges. Firstly by setting $\delta j_{++}=\delta \varsigma_{+-}=0$ (which no longer constrains the solution space). Secondly, performing a field-dependent redefinition of the generators of residual symmetries $\epsilon$ and $\sigma$.

In this paper, we take the first route. Hence, the charges reduce to
{\begin{align}
{\ndelta Q_{\underline{\sigma}}}&={-}\dfrac{4\sqrt{42}(2\Xi^2L^2-17)\varsigma_{+-}}{21\Xi^2L^3\sqrt{\Xi^2L^2+2}j_{++}}\int_{0}^{2\pi} d\phi \sigma(x^{+})\delta h, \\
{\ndelta Q_{\underline{\epsilon}}}&=\dfrac{2\sqrt{42}}{21\sqrt{\Xi^2L^2+2}\Xi^2 L^3(2\Xi^2 L^2+25)j_{++}}\times\nonumber\\
&\int_{0}^{2\pi}d\phi\; \epsilon(x^{+}) \left[16 j_{++}(\Xi^2L^2+2)^{2}\delta f_{++}-3(2\Xi^2L^2-17)(2\Xi^2L^2+11) h\delta h\right].
\end{align}
Now, by integrating the charges, we have
{\begin{align}\label{eqq43}
{ Q_{\underline{\sigma}}}&={-}\dfrac{4\sqrt{42}(2\Xi^2L^2-17)\varsigma_{+-}}{21\Xi^2L^3\sqrt{\Xi^2L^2+2}j_{++}}\int_{0}^{2\pi} d\phi \sigma(x^{+})(h+h_{0}), \\
{ Q_{\underline{\epsilon}}}&=\dfrac{\sqrt{42}}{21\sqrt{\Xi^2L^2+2}\Xi^2 L^3(2\Xi^2 L^2+25)j_{++}}\times\nonumber\\
&\int_{0}^{2\pi}d\phi\; \epsilon(x^{+}) \left[32 j_{++}(\Xi^2L^2+2)^{2} f_{++}-3(2\Xi^2L^2-17)(2\Xi^2L^2+11) h^{2}\right].
\end{align}
In \eqref{eqq43}, we fix $h_{0}$ by demanding that the charge associated with the exact isometries of (\ref{metricADM}) matching with \eqref{eqmassang}. Therefore, we get
{\begin{equation}\label{eqqh0}
h_{0}=\dfrac{2+\Xi^2 L^2}{42}.
\end{equation}}
Now the charge algebra of $u(1)$ sector is
\begin{equation}
\delta_{\sigma_{2}} Q_{\sigma_{1}}[g]=Q_{[\sigma_{1},\sigma_{2}]}+K_{\sigma_{1},\sigma_{2}}.
\end{equation}
Since the vectors are field independent,  therefore $Q_{[\sigma_{1},\sigma_{2}]}=0$, the central extension for the $u(1)$ sector is
\begin{equation}
K_{\sigma_{1},\sigma_{2}}={-}\dfrac{8\sqrt{42}(2\Xi^2L^2-17)\varsigma_{+-}^{2}}{21\Xi^2L^3\sqrt{\Xi^2L^2+2}j_{++}}\int_{0}^{2\pi} d\phi \sigma_{1}\sigma^{\prime}_{2}.
\end{equation}
Using the mode decomposition $\sigma_{1}=e^{imx^{+}}$, $\sigma_{2}=e^{inx^{+}}$, and calling $ Q_{\underline{\sigma}^{1}}=P_{m} $, $ Q_{\underline{\sigma}^{2}}=P_{n} $, we obtain
\begin{equation}
i\left\lbrace P_{m},P_{n} \right\rbrace =m\dfrac{k}{2}\delta_{m+n,0}\;\;\;\;\text{with}\;\;\;\;k=\dfrac{{16}\pi\sqrt{42}(2\Xi^2L^2-17)\varsigma_{+-}^{2}}{21\Xi^2L^3\sqrt{\Xi^2L^2+2}j_{++}}.
\end{equation}
This is a centrally extended $u(1)$ algebra with central extension $k$-colloquially called the Kac-Moody level. For the Virasoro sector using the modified Lie bracket, we have
\begin{align}
&\left\lbrace Q_{\underline{\epsilon}^{1}},Q_{\underline{\epsilon}^{2}} \right\rbrace =\delta_{\epsilon_{2}} Q_{\epsilon_{1}}[g]=\dfrac{\sqrt{42}}{21\sqrt{\Xi^2L^2+2}\Xi^2 L^3(2\Xi^2 L^2+25)j_{++}}\int_{0}^{2\pi}d\phi \left(\epsilon_{1}\epsilon_{2}^{\prime}-\epsilon_{2}\epsilon_{1}^{\prime}\right)\times\nonumber\\
&\left[32 j_{++}(\Xi^2L^2+2)^{2} f_{++}-3(2\Xi^2L^2-17)(2\Xi^2L^2+11) h^{2}\right]-\dfrac{16\pi \sqrt{42}}{(21)^{2}}\dfrac{(\Xi^2L^2+2)^{\frac{3}{2}}}{\Xi^2 L}\int_{0}^{2\pi}d\phi \epsilon_{1}\epsilon_{2}^{\prime\prime\prime}.
\end{align}
Using the mode decomposition representation $\epsilon_{1}=e^{imx^{+}}$, $\epsilon_{2}=e^{inx^{+}}$, and calling $Q_{\underline{\epsilon}_{1}}=L_{m}$, $Q_{\underline{\epsilon}_{2}}=L_{n}$, one can obtain the following
\begin{equation}
i\left\lbrace L_{m}, L_{n} \right\rbrace =(m-n)L_{m+n}+\dfrac{c}{12}m^{3}\delta_{m+n,0},\;\;\;\; c=\dfrac{16\pi}{7}\sqrt{\dfrac{2}{21}}\dfrac{(\Xi^2 L^2+2)^{\frac{3}{2}}}{\Xi^2 L}.
\end{equation}
Finally, the algebra is
\begin{align}
i\left\lbrace L_{m}, L_{n} \right\rbrace &=(m-n)L_{m+n}+\dfrac{c}{12}m^{3}\delta_{m+n,0},\\
i\left\lbrace L_{m}, P_{n} \right\rbrace &=-nP_{n+m},\\
i\left\lbrace P_{m},P_{n} \right\rbrace &=m\dfrac{k}{2}\delta_{m+n,0},
\end{align}
with the central extensions
\begin{equation}\label{centralex}
c=\dfrac{16\pi}{7}\sqrt{\dfrac{2}{21}}\dfrac{(\Xi^2 L^2+2)^{\frac{3}{2}}}{\Xi^2 L},\;\;\;\;\;k=\dfrac{{16}\pi\sqrt{42}(2\Xi^2L^2-17)\varsigma_{+-}^{2}}{21\Xi^2L^3\sqrt{\Xi^2L^2+2}j_{++}}.
\end{equation}
As can be seen, $c$ is independent of the solution space ($j_{++},\varsigma_{+-}$) whereas $k$ depends on them. Now let us study the null warped limit by taking $2\Xi^2 L^2=17$. In this case, the $u(1)$ level and charges vanish identically ($k=0$), and we are left with a Virasoro symmetry algebra with a central extension
\begin{equation}
c=\dfrac{48\pi L}{17},\;\;\;\; \Xi L \to \sqrt{\dfrac{17}{2}}.
\end{equation}
Moreover, as discussed earlier, the CSS limit is reached by setting $2\Xi^2L^2=17$ and $j_{++}=0$ while keeping $\frac{2\Xi^2L^2-17}{j_{++}}=\Delta$ constant. The charges read as follows.
\begin{align}
{ Q_{\underline{\sigma}}}=\dfrac{16}{357}\dfrac{\Delta \varsigma_{+-}}{L}\int_{0}^{2\pi} d\phi \sigma(x^{+})(h+h_{0}),\;\;\;\;\; { Q_{\underline{\epsilon}}}&=\dfrac{8}{17L}\int_{0}^{2\pi}d\phi\epsilon(x^{+}) \left[2f_{++}-\dfrac{\Delta h^2}{21}\right],
\end{align}
while the central extensions become
\begin{equation}
c=\dfrac{48\pi L}{17},\;\;\;\;k=-\dfrac{32\pi \Delta \varsigma_{+-}^{2}}{357L}.
\end{equation}
This limit coincides with our results in \cite{Setare:2021ugr}.
We now turn our attention to the WBTZ black holes. These are reached by restricting the solution space to
\begin{align}
j_{++}&=\dfrac{2\Xi^2 L^2-17}{168GL(ML-J)},\;\;\;\; \varsigma_{+-}={-}\dfrac{2}{21}(2+\Xi^2 L^2),\;\;\;\;\gamma_{--}=\dfrac{4GL(2+\Xi^2 L^2)(LM-J)}{21},\\f_{++}(x^{+})&=\dfrac{GL (ML+J)(25+2\Xi^2 L^2)}{21},\;\;\;
h(x^{+})=0.
\end{align}
Therefore, their charges in the quadratic ensemble take the form
\begin{align}
{P_{m}}&={\dfrac{128\pi \sqrt{42}}{21}\left(ML-J\right)\dfrac{h_{0}\sqrt{2+\Xi^2L^2}}{\Xi^2 L^2}\delta_{m,0},}\\
L_{m}&={\dfrac{64\pi\sqrt{42}}{(21)^{2}}\dfrac{(\Xi^2L^2+2)^{\frac{3}{2}}}{\Xi^2L^2}\left(ML+J\right)\delta_{m,0}}
\end{align}
with $M$ and $J$ are the global charges from the Einstein's gravity sector. The NMG mass and angular momentum of these solutions are defined as
\begin{equation}
\mathcal{M}=Q_{\partial_{t}}=\dfrac{1}{L}(Q_{\partial_{+}}+Q_{\partial_{-}}),\;\;\;\; \mathcal{J}=Q_{\partial_{\phi}}=Q_{\partial_{+}}-Q_{\partial_{-}},
\end{equation}
and we also have
\begin{equation}
Q_{\partial_{-}}=P_{0},\;\;\;\;Q_{\partial_{+}}=L_{0}.
\end{equation}
Then, we obtain the relationship between the NMG mass and angular momentum and the zero modes of the charges
{\begin{align}\label{eqmassang}
\mathcal{M}&=\dfrac{1}{L}(P_{0}+L_{0})=\dfrac{128\pi\sqrt{42}(2+\Xi^{2}L^{2})^{\frac{3}{2}}M}{441\Xi^2 L^3},\nonumber\\
\mathcal{J}&=L_{0}-P_{0}=\dfrac{128\pi\sqrt{42}(2+\Xi^{2}L^{2})^{\frac{3}{2}}J}{441\Xi^2 L^3}.
\end{align}}
Here, we explicitly show that the bulk solution space has a symmetry algebra that can be identified with that of a WCFT in the quadratic ensemble. 

\section{Entropy matching}\label{secc4}
To study the thermodynamics of WBTZ solutions, we bring the line element to the ADM form \cite{Aggarwal:2020igb,Kraus:2005zm}, 
\begin{equation}\label{metricADM}
ds^{2}=-N(r)^{2}dt^2+\dfrac{dr^2}{f(r)^2}+R(r)^2(N^{\phi}(r)dt+d\phi)^{2},
\end{equation}
with
\begin{align}
N^2(r)&=-\dfrac{4(2H^2-1)(J-ML)(16J^2L^2-8ML^2r^2+r^4)}{L(16H^2L^2J^2+H^2r^4-4Lr^2(ML+J(2H^2-1)))},\\
f^2(r)&=\dfrac{16J^2}{r^2}-8M+\dfrac{r^2}{L^2},\\
R^2(r)&=-\dfrac{16H^2J^2L^2+H^2r^4-4Lr^2(2H^2J-J+ML)}{4L(ML-J)},\\
N^{\phi}(r)&=\dfrac{H^2r^4-8MH^2L^2r^2-16JL^2(J(H^2-1)+LM(1-2H^2))}{L(H^2r^4+16H^2J^2L^2-4Lr^2(ML+J(2H^2-1)))},
\end{align}
where $H$ can be found in \eqref{lasteq}. Here, we follow Wald's approach to compute the entropy as follows \cite{Wald1,Wald2}
\begin{equation}
S=-2\pi \int_{\Sigma}d\phi\sqrt{\eta}\dfrac{\delta L}{\delta \mathcal{R}_{\mu \nu \alpha \beta}}\epsilon_{\mu \nu}\epsilon_{\alpha \beta},
\end{equation}
where $L$ is the Lagrangian, $\epsilon_{\mu \nu}=-2\sqrt{-\zeta}\delta^{t}_{[\mu}\delta^{r}_{\nu]}$, $\eta$ is a determinant of the induced metric on the horizon and $\zeta$ is the determinant of the bifurcate metric. For the NMG action, we have
\begin{small}
\begin{align}
\dfrac{\delta L}{\delta \mathcal{R}_{\mu \alpha \beta \nu}}=&\left(\dfrac{1}{2}+\dfrac{3}{8\Xi^2} R\right)\left(g^
{\mu \beta}g^{\alpha \nu}-g^{\mu \nu}g^{\alpha \beta}\right)-\dfrac{1}{2\Xi^{2}}\left(\mathcal{R}^{\mu \beta}g^{\alpha \nu}-\mathcal{R}^{\alpha \beta}g^{\mu \nu}-\mathcal{R}^{\mu \nu}g^{\alpha \beta}+\mathcal{R}^{\alpha \nu}g^{\mu \beta}\right).
\end{align}
\end{small}
For given line element \eqref{metricADM}, the entropy of a stationary black hole in NMG is
 \begin{align}
 S^{NMG}_{\pm}=&\pi^{2}R(r_{\pm})-\dfrac{\pi^2 f(r)^{2}}{8\Xi^{2}N(r)^{4}}[-4f^{2}NRN^{\prime\prime}+5f^2R^3N^{\phi \prime 2}-4ff^{\prime}NN^{\prime}R+4f^{2}N^{2}R^{\prime\prime}\nonumber\\
 &+4ff^{\prime}N^{2}R^{\prime}+4f^{2}NR^{\prime}N^{\prime}]_{r_{\pm}},
 \end{align}
where $r_{\pm}$ are the horizons of black holes i.e., $f(r_{\pm})=0$, given by
\begin{equation}
r_{\pm}=2\sqrt{GL}\sqrt{LM\pm\sqrt{L^2M^2-J^2}}.
\end{equation}
For the WBTZ black hole, the entropy at the outer horizon ($S_+$) is
\begin{equation}\label{eqent}
{S_{+}=\dfrac{64\pi^2 (1-2H^2)^{\frac{3}{2}}r_{+}}{-17+42H^2}}.
\end{equation}
The Hawking temperature and angular velocity of the black hole are given as
\begin{align}
T &=\dfrac{r_{+}^{2}-r_{-}^{2}}{2\pi r_{+}L^2}=\dfrac{2\sqrt{L^2M^2-J^2}}{\pi L^{\frac{3}{2}}\sqrt{ML+\sqrt{M^2L^2-J^2}}}, \\
\Omega&=\dfrac{r_{-}}{Lr_{+}}=\dfrac{\sqrt{ML-\sqrt{L^2M^2-J^2}}}{L\sqrt{ML+\sqrt{M^2L^2-J^2}}}. \label{eqangu}
\end{align}
As expected, the thermodynamic quantities ($\mathcal{M,J},S,T,\Omega$) satisfy the first law of black hole mechanics and the Smarr formula as follows
\begin{equation}
d\mathcal{M}=TdS+\Omega d\mathcal{J},\;\;\;M=\dfrac{1}{2}TS+\Omega J.
\end{equation}
{As a comparison, we shall compute the entropy of the WBTZ black hole by implementing the Cardy formula. First, we define the vacuum of the theory by promoting the local symmetries to global symmetries. The killing vectors of the WBTZ metric, in addition to $\partial_{t}$ and $ \partial_{\phi} $, are given by}
\begin{align}
\Upsilon^{r}_{\pm}=&e^{\pm 2\sqrt{\frac{2G(J+LM)}{L}}\left(\frac{t}{L}+\phi\right)}\frac{\sqrt{16G^{2}J^{2}L^2-8GL^{2}Mr^{2}+r^{4}}}{2r},\\
\Upsilon^{t}_{\pm}=&\mp e^{\pm 2\sqrt{\frac{2G(J+LM)}{L}}\left(\frac{t}{L}+\phi\right)}\dfrac{L(4GJL+r^{2})}{4\sqrt{\frac{2G(J+ML)(16G^{2}J^{2}L^{2}-8GL^{2}Mr^{2}+r^{4})}{L}}},\\
\Upsilon^{\phi}_{\pm}=&\pm e^{\pm 2\sqrt{\frac{2G(J+LM)}{L}}\left(\frac{t}{L}+\phi\right)}\frac{4GL(J+2ML)-r^{2}}{4\sqrt{\frac{2G(J+ML)(16G^{2}J^{2}L^{2}-8GL^{2}Mr^{2}+r^{4})}{L}}}.
\end{align}
These local isometries are globally extended if the exponential is $2\pi$ periodic in $\phi$. Thus, the WBTZ solution admits global isometries if and only if
\begin{equation}
2\sqrt{\dfrac{2G(J+ML)}{L}}=\pm i.
\end{equation}
This equation, therefore, has a solution $M=-1/8G-J/L$. Therefore, the line element of a particular vacuum is obtained by setting $J=0, M=-1/8G$ in \eqref{metricADM}. Thus, the vacuum line element is 
\begin{align}
ds_{vac}^{2}&=(L^2+r^2)(2H^2r^2+L^2(2H^2-1))\dfrac{dt^2}{L^4}+\dfrac{L^2dr^2}{\left(L^2+r^2\right)}+4H^2r^2(L^2+r^2)\dfrac{dtd\phi}{L^3} \nonumber \\
&~~~~+\left(r^2+\dfrac{2H^2r^4}{L^2}\right)d\phi^2.
\end{align}
Remark that, the vacuum metric above reduces to global AdS$_3$ when $H=0$. From the metric above, the NMG charges are given by
{\begin{align}
\mathcal{M}&=-\dfrac{8\pi\sqrt{1-2H^{2}}}{17-42H^2}\left[1-2H^2+{4h_{0}}\right],\\
\mathcal{J}&=-\dfrac{8\pi L\sqrt{1-2H^{2}}}{17-42H^2}\left[1-2H^2-{4h_{0}}\right].
\end{align}}
In the quadratic ensemble, the Warped Cardy formula takes the form
\begin{equation}\label{eeqq}
S_{WCFT}=4\pi\sqrt{-P_{0}^{vac}P_{0}}+4\pi\sqrt{-L_{0}^{vac}L_{0}}.
\end{equation}
The zero modes for the vacuum metric become
{\begin{equation}
P_{0}^{vac}=-{32\pi h_{0}L}\dfrac{\sqrt{1-2H^2}}{17-42H^2},\;\;\;L_{0}^{vac}=-{16\pi L}\dfrac{(1-2H^2)^{\frac{3}{2}}}{17-42H^2}.
\end{equation}}
Inserting these in \eqref{eeqq} and using \eqref{eqqh0}, one can find
{\begin{equation}
S_{WCFT}=\dfrac{64\pi^2\sqrt{2L}(1-2H^2)^{\frac{3}{2}}}{17-42H^{2}}\left[\sqrt{ML-J}+\sqrt{ML+J}\right]
\end{equation}}
This expression matches the bulk thermodynamic entroy of the WBTZ
\eqref{eqent}. 
\section{Conclusion}\label{secc5}
In this work, using the field truncation \eqref{boundcond} (inspired by the WBTZ black hole), we obtain the solution space, the residual asymptotic symmetry, and the surface charges in the framework of new massive gravity. The solution space constructed of two arbitrary chiral functions ($h(x^{+}), f_{++}(x^{+})$) and two constants ($j_{++}, \varsigma_{+-}$).
The residual asymptotic symmetry group is constituted of the direct sum of a Witt algebra and a $u(1)$ algebra. The surface charges for the asymptotic symmetries have been obtained. We find that the charges are finite, but non-integrable. The charges become integrable when considering $\delta j_{++}=\delta \varsigma_{+-}=0$ on the solution space. The integrable charges group, similar to the residual symmetry group, is constituted of $ Vir \otimes u(1)$ Kac-Moody algebra 
with the central charges given in \eqref{centralex}. 
Finally, we obtain the entropy of the WBTZ black hole using the Cardy formula, demonstrating that the Cardy formula is applicable in the quadratic ensemble for WCFTs. 

It would be interesting to extend the domain of validity of these fall-offs for the other 3D massive gravity theories (such as GMMG,  EGMG) and different gauges (such as Bondi and Bondi-Weyl gauge). Also, it is interesting to compute the linearized energy excitations (energy of gravitons) in WAdS$_3$. We leave these works for the future.

\section*{Acknowledgements}
This research has received funding support from the NSRF via the Program Management Unit for Human Resource and Institutional Development, Research and Innovation grant number $B13F680083$.

\appendix
\section{WBTZ metric}\label{app:A}
The WBTZ metric can be obtained by a deformation of the BTZ black hole spacetime as follows \cite{Israel:2004vv}
\begin{equation}\label{WBTZsol}
ds_{WBTZ}^{2}=ds_{BTZ}^{2}-2H^2\xi \otimes\xi,
\end{equation}
where
\begin{equation}
ds_{BTZ}^{2}=\dfrac{L^2r^2dr^2}{16J^2L^2-8ML^2r^2+r^4}+(4ML^2-r^2)dx^{+}dx^{-}+2L(LM+J)dx^{+2}+2L(LM-J)dx^{-2},
\end{equation}
and
\begin{equation}
\xi=-\dfrac{1}{\sqrt{2GL(LM-J)}}\partial_{-}.
\end{equation}
This metric is a solution for the NMG field equation under the conditions
\begin{equation}
\lambda=\dfrac{\Xi^4 L^4-32\Xi^2 L^2+16}{21\Xi^2 L^4},\;\;\;\;H^2=\dfrac{-2\Xi^2 L^2+17}{42}. \label{lasteq}
\end{equation}
For $2\Xi^2 L^2=17$, $H$ becomes zero and the metric WBTZ becomes the BTZ metric.
 
\end{document}